\documentclass[prd,aps,twocolumn]{revtex4}
\usepackage{graphicx} 

\begin{document}
\title{Hadronic nature of high-energy emission from the Galactic Ridge}
\author{A.~Neronov$^{1,2}$, D.~Semikoz$^{1}$, J.~Aublin$^{1}$, M.~Lamoureux$^{3}$, A.~Kouchner$^{1}$}
\affiliation{$^1$ Astroparticules et Cosmologie, Université Paris Cité, CNRS, Astroparticule et Cosmologie,  F-75006 Paris, France\\
$^2$Laboratory of Astrophysics, Ecole Polytechnique Federale de Lausanne, 1015, Lausanne, Switzerland \\
$^{3}$ Centre for Cosmology, Particle Physics and Phenomenology -- CP3, Université Catholique de Louvain, Louvain-la-Neuve, Belgium}

\begin{abstract}
    We show that the IceCube observation of the Galactic neutrino flux component confirms the hint of detection of neutrinos from the Galactic Ridge (the inner part of the Milky Way disk within the Galactic longitude $|l|<30^\circ$), previously reported by the ANTARES collaboration. This confirmation indicates that the bulk of the high-energy flux from the Galactic Ridge in multi-TeV band is produced by interactions of high-energy protons and atomic nuclei, rather than electrons. We show that both ANTARES and IceCube measurements  agree with the Fermi-LAT telescope measurements of the gamma-ray emission from the Ridge.  The multi-messenger (neutrino plus gamma-ray) spectrum of the Ridge over a broad energy range from 10 GeV to 10 TeV is consistent with a model of pion decay emission produced by a power-law distribution of protons with a slope $\Gamma\simeq 2.5$, harder than that of the locally observed cosmic ray spectrum. This provides for the first time an unambiguous multi-messenger demonstration of the variability of the spectrum of cosmic rays across the Galactic disk.
\end{abstract}

\maketitle

\section{Introduction.} 

IceCube collaboration has recently reported the detection of neutrino signal from the Milky Way \cite{icecube_gal} in the ``cascade'' type neutrino-induced events in the detector. Evidence for the presence of the Galactic component in the astrophysical neutrino flux in the same detection channel has been previously seen in an earlier data release of IceCube \cite{2016APh....75...60N}. Refs. \cite{icecube_gal} and \cite{2016APh....75...60N} have used different approaches for the Galactic neutrino flux search: fitting of a theoretically-motivated  all-sky map template \cite{icecube_gal} vs. a sky-model-independent approach of searching for the excess signal at low Galactic latitudes \cite{2016APh....75...60N}. A different approach has been considered by the ANTARES collaboration that has concentrated on the search for a neutrino signal from the Galactic Ridge, the central part of the Milky Way disk within Galactic longitude $|l|<30^\circ$. An excess of neutrino events from the Ridge direction has been reported \cite{2023PhLB..84137951A}. This excess is consistent with the initial estimates of the neutrino flux from the Ridge derived from the IceCube data \cite{2014PhRvD..89j3002N}.

The motivation for the search for a neutrino signal from the Galactic Ridge stems from the fact that the Ridge projection on the sky includes the Galactic Bar and the innermost parts of the spiral arms. Tracers of star formation including pulsars and supernovae indicate that the star formation rate (and hence cosmic ray injection rate) in the Milky Way disk peaks at the distance of about $d_{\rm sfr}\simeq 4$~kpc from the center \cite{2012ApJ...750....3A}. Given that the distance of the Sun from the Milky Way center is $d_\odot\simeq 8$~kpc, one expects the region of the most intense star formation to span the range $|l|\le \arcsin(d_{\rm sfr}/d_\odot)\simeq 30^\circ$ of the Galactic longitude.

The Galactic Ridge is a bright $\gamma$-ray source that hosts prominent isolated sources superimposed on the diffuse emission that has been detected up to TeV energy range by H.E.S.S. telescope \cite{2014PhRvD..90l2007A} and Fermi-LAT \cite{2020A&A...633A..94N}. It is not clear a priori if this diffuse TeV emission is predominantly produced by interactions of cosmic ray protons and nuclei in the interstellar medium or if it originates from interactions of cosmic ray electrons (Compton scattering of starlight). If the diffuse flux is produced by the interactions of cosmic ray protons, its peculiar spectral properties suggest that the spectrum of cosmic rays in the inner part of the Milky Way is different from that observed locally \cite{2000A&A...362..937A,2015arXiv150507601N,2016PhRvD..93l3007Y}.

A definitive answer to the question of whether the $\gamma$-ray emission from the Galactic Ridge is of ``hadronic'' or ``leptonic'' origin can be found if the neutrino signal from the Ridge direction is measured. This is because only the hadronic emission has a neutrino counterpart. It is generated through the production and decay of pions, with neutral pions decaying directly into $\gamma$-rays and charged pions producing neutrinos. If the parent proton and nuclei cosmic ray spectrum is a power-law with the slope $\Gamma$, the pion decay $\gamma$-ray and neutrino spectra are expected to be nearly power-laws in the multi-TeV range with similar slopes and with nearly identical normalizations \cite{2006ApJ...647..692K,2006PhRvD..74c4018K,2019CoPhC.24506846K}. 

In what follows, we show that the IceCube detection of the Galactic component of the neutrino flux in the multi-TeV band provides a rather precise measurement of the Galactic Ridge neutrino signal. It confirms the Ridge neutrino flux estimate of ANTARES and resolves the problem of the nature of the multi-TeV emission from the Ridge, confirming its "hadronic" origin.

\section{Neutrino flux from the Galactic Ridge derived from the IceCube measurement.} 

The IceCube Galactic Plane signal \cite{icecube_gal} has been found by fitting a pre-defined model template to the all-sky data. This model template, derived from the $\gamma$-ray data of Fermi-LAT, is in fact dominated by the flux from the Galactic Ridge.
This sky region provides the dominant contribution to the test statistic of the likelihood analysis performed in \cite{icecube_gal}. Thus, the model fitting mostly constrains the flux from the Galactic Ridge direction, while the flux from other parts of the sky e.g., the outer part of the Galactic Plane, or the flux from high Galactic latitude regions cannot really be constrained, because they provide only minor contribution to the test statistic. 

\begin{figure}
\includegraphics[width=\linewidth]{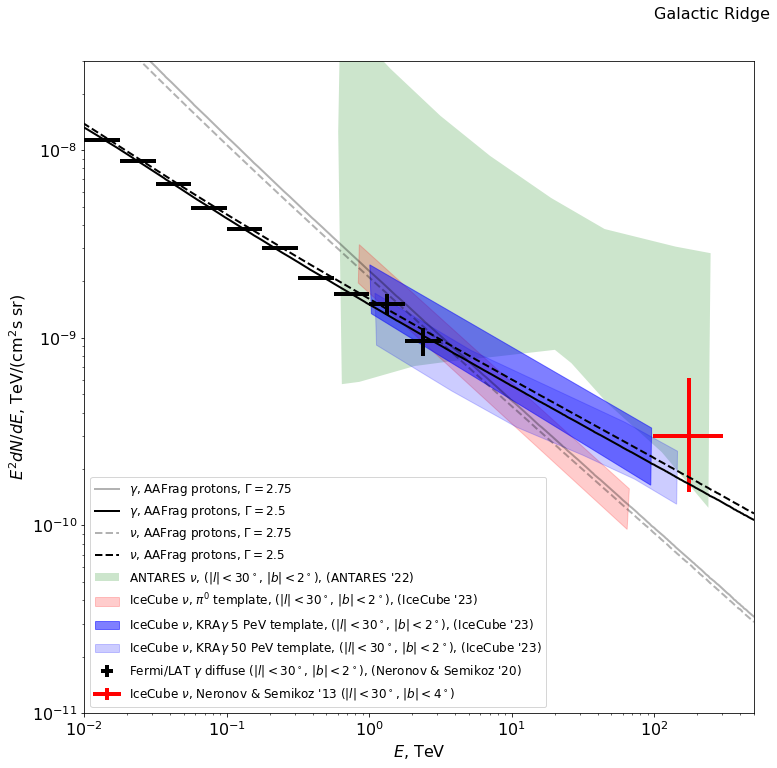}
\caption{Multi-messenger spectrum of the diffuse emission from the Galactic Ridge. Fermi-LAT $\gamma$-ray data are from \cite{2020A&A...633A..94N}. IceCube neutrino flux estimate shown by the red data points is from \cite{2014PhRvD..89j3002N}. ANTARES hint of the signal from the Galactic Ridge is from \cite{2023PhLB..84137951A}. Lighter and darker blue shaded regions show the estimates of the IceCube signal from the direction of the Ridge done following the method of Ref. \cite{icecube_gal} assuming the KRA$\gamma$ all-sky model templates with cut-offs at 5 PeV and 50 PeV. Red shaded region shows the estimate of the Galactic Ridge flux derived using the $\pi^0$ all-sky model template of \cite{icecube_gal}. Black (grey) curves show the $\gamma$-ray (solid) and neutrino (dashed) pion decay spectra calculated for power-law distributions of protons with slope $\Gamma=2.5$ ($\Gamma=2.75$), calculated using AAFrag code \cite{2019CoPhC.24506846K,2021PhRvD.104l3027K}. All neutrino fluxes correspond to the all-flavor $\nu+\bar\nu$ emission with $1\sigma$ uncertainty bands.}
\label{fig:spectrum}
\end{figure}

To estimate the IceCube neutrino flux from the direction of the Ridge $-30^\circ<l<30^\circ$, we use the same approach as in the Ref. \cite{icecube_gal}, where the flux from the Galactic Plane in the Galactic longitude range $25^\circ<l<100^\circ$ and $50^\circ<l<200^\circ$ was estimated. We use the all-sky model template maps provided in the Supplementary material of Ref. \cite{icecube_gal} to find the fraction of the total flux that comes from the Ridge and re-scale the all-sky total flux estimate derived in \cite{icecube_gal} to get an estimate of the flux from the Ridge. The result is shown by the darker and lighter blue shading for the IceCube best fit of the KRA$\gamma$ models \cite{2015ApJ...815L..25G} with cut-offs at $5$~PeV and 50~PeV. The two estimates agree with each other in a wide energy range from 1 to 100~TeV. The difference between the two estimates can be considered as a systematic modeling uncertainty. Further assessment of the model uncertainty can be done considering the $\pi^0$ all-sky model template of Ref. \cite{icecube_gal} instead of the KRA$\gamma$ template. The estimate of the neutrino flux from the Ridge using this model template is shown by the red shading in Fig. \ref{fig:spectrum}. One can see also the $\pi^0$ template model flux estimate agrees with the KRA$\gamma$ model flux estimates in the energy range 1-10~TeV. However, the slopes of the spectra of the Ridge computed under different all-sky model assumptions cannot agree "by design": the $\pi^0$ model assumes that the slope of the neutrino emission spectrum is the same all over the sky (close to the slope of the parent proton spectrum, $\Gamma=2.7$), while the versions of the KRA$\gamma$ model used in the IceCube analysis of Ref. \cite{icecube_gal} assumed constant "sky-averaged" slope close to $2.5$.

The  IceCube Galactic Ridge neutrino flux estimate is consistent with the previous report of the hint of the neutrino signal from the Galactic Ridge by the ANTARES Collaboration  \cite{2023PhLB..84137951A} shown by the green shading in Fig. \ref{fig:spectrum}. It is also in agreement with the estimate of Ref. \cite{2014PhRvD..89j3002N} for the neutrino flux from the Ridge based on early IceCube measurements of the astrophysical neutrino flux in High-Energy Starting Events (HESE) channel shown by the red data point in Fig. \ref{fig:spectrum}.

The neutrino flux from pion decays should always have a $\gamma$-ray counterpart that has spectral characteristics similar to those of the neutrino signal (see model lines in Fig. \ref{fig:spectrum}). Thus, the neutrino flux from the hadronic source should be comparable to (or smaller than) the $\gamma$-ray flux, because the $\gamma$-ray flux can have an additional "leptonic" component that does not have a neutrino counterpart. From Fig. \ref{fig:spectrum} one can see that the neutrino flux estimates from both IceCube and ANTARES in the TeV energy range agree well with the $\gamma$-ray flux measurements by Fermi-LAT from the same sky region in the same energy range. 

\section{Implications for the cosmic ray spectrum in the Milky Way.} 

The consistency of the IceCube and ANTARES neutrino flux estimates from the Galactic Ridge with the measurements of the diffuse $\gamma$-ray emission from the Ridge in the TeV energy range  with Fermi-LAT shows that the bulk of the multi-messenger emission from this part of the sky originates from the decays of pions produced in interactions of high-energy protons and atomic nuclei, rather than from electron interactions. This is the first time when such an unambiguous distinction between "hadronic" and "leptonic" emission mechanisms can be done.

Multi-messenger measurements of high-energy emission from the Galactic Ridge region also allow us to establish a fact important for the physics of cosmic rays. It shows that the locally observed slope of the cosmic ray spectrum is not representative of the average cosmic ray spectrum in the Milky Way. Instead, the cosmic ray spectrum varies depending on the position in the Galactic disk. The multi-messenger data provide for the first time a measurement of the cosmic ray spectrum in the Galactic Ridge region extending up to about $4$~kpc distance from the Galactic Center. Solid and dashed black model lines in Fig. \ref{fig:spectrum} show model fits to the multi-messenger Galactic Ridge spectrum. The solid line corresponds to the $\gamma$-ray emission from the decays of neutral pion decays produced in interactions of high-energy protons with the power-law spectrum with the slope $\Gamma=2.5$. One can see that such a model describes the $\gamma$-ray data on the diffuse emission from the Ridge, over a broad energy range from 10~GeV up to 3 TeV. The dashed line shows the all-flavor neutrino flux from the decays of charged pions produced in interactions of the same protons. One can see that this model also correctly describes the neutrino data in the multi-TeV energy range. 

Grey solid and dashed lines in Fig. \ref{fig:spectrum} show the model of pion decay emission from cosmic rays distributed according to a power-law with the slope $\Gamma=2.75$, close to that of the locally observed cosmic ray spectrum slope. One can see that if the model neutrino spectrum is normalized to the IceCube estimate of the neutrino flux, the gamma-ray model spectrum over-predicts the  Fermi-LAT $\gamma$-ray flux measurements at nearly all energies below TeV. Thus, the possibility that the spectrum of cosmic rays in the Galactic Ridge has the slope close to that of the locally measured cosmic ray spectrum is ruled out. 

We notice that the same problem is encountered if the all-sky Galactic neutrino spectrum reported in Ref. \cite{icecube_gal} is considered. The all-sky flux estimate obtained using the $\pi^0$ template with the assumed slope $\Gamma\simeq 2.7$  \cite{icecube_gal} largely over-predicts the diffuse all-sky $\gamma$-ray flux measured by Fermi-LAT \cite{2012ApJ...750....3A}, and even over-predicts the total (diffuse plus resolved sources) all-sky $\gamma$-ray flux  \cite{2018PhRvD..98b3004N} by at least a factor-of-two in the GeV band. 

The all-sky neutrino flux estimates obtained in \cite{icecube_gal}  using both KRA$\gamma$ templates with the sky-averaged slope $\Gamma\simeq 2.5$  predict the GeV band pion decay $\gamma$-ray flux that is an order-of-magnitude below the $\gamma$-ray diffuse emission flux measured by Fermi-LAT in the GeV range \cite{2012ApJ...750....3A}. At the same time, Ref. \cite{2012ApJ...750....3A} analysis suggests that about half of the all-sky diffuse emission flux in the GeV range originates from the neutral pion decays. 

These inconsistencies between the all-sky neutrino and $\gamma$-ray pion decay diffuse flux measurements provide additional confirmation of the variability of the cosmic ray spectrum across the Milky Way: models with the position-independent slope of the cosmic rays spectrum all across the Milky Way do not describe the all-sky multi-messenger data.  

\section{Discussion.} 

Changes of the slope of the cosmic ray spectrum at different locations in the Milky Way galaxy  can be due to several mechanisms. The properties of the cosmic ray population in the Galactic halo are determined by the balance between the cosmic ray injection rates from sources and the escape of cosmic rays from the halo. If cosmic rays are injected with an average power-law spectrum with the slope $\Gamma_{\rm inj}$, while the escape time of cosmic rays from the halo depends on energy as $t_{\rm esc}\propto E^{-\delta}$, the steady-state cosmic ray spectrum in the halo is expected to be a power-law with the slope $\Gamma=\Gamma_{\rm inj}+\delta$. The observation of variability of $\Gamma$ across the Galactic disk may suggest that $\Gamma_{\rm inj}$ and/or $\delta$ experience spatial variations. For example, $\Gamma_{\rm inj}$, the slope of the injection spectrum averaged over source population(s), may change as a function of the rate of occurrence of sources of different types, or with the typical source environment in different parts of the Galactic Disk. Otherwise, cosmic rays escape from the Galactic halo through diffusion in the Galactic magnetic field. The exponent of the energy dependence of the escape time, $\delta$ may depend on the structure of the turbulent component of the Galactic magnetic field, on the relative strength of this turbulent component compared to the regular field component, on the geometry of the regular field component \cite{2018JCAP...07..051G}. Both the average characteristics of the cosmic ray source populations and the structure of the Galactic magnetic field change with position in the Galactic Disk \cite{2012ApJ...757...14J}. In this sense, it is not surprising that the cosmic ray spectrum in the Milky Way is not "universal", but varies across the Galaxy. Finally, it is possible that the locally observed cosmic ray spectrum is "peculiar", in the sense that it is influenced by a specific local source, or by a peculiarity of the history of star formation in the Solar neighborhood \cite{2015PhRvL.115r1103K,2013ApJ...769..138E}. In such a model, even though the "average" Galactic cosmic ray spectrum may be well defined, any local measurement of the cosmic ray spectrum at a fixed location may not measure it, because the spectrum fluctuates from point to point. Modeling of such a situation was recently performed in Refs. \cite{Bouyahiaoui:2020rkf,Bouyahiaoui:2021rev,Giacinti:2023upw}. A detailed discussion of galactic CR models was done in a recent review \cite{Kachelriess:2019oqu}.

It will be possible to distinguish between the three possibilities described above when multi-messenger measurements of the spectra of hadronic emission from different parts of the Milky Way disk will become available (the spectrum of hadronic emission from the Galactic Ridge reported here is just the first example of such a measurement). The new generation of neutrino telescope KM3NeT \cite{2016JPhG...43h4001A} will provide additional and independent data with good energy and angular resolution, that will help to discriminate between Galactic cosmic ray models.

\section{Conclusions.} 

In this work, we have derived a measurement of the neutrino flux from the Galactic Ridge from IceCube detection of the Galactic astrophysical neutrino flux. We have shown that the IceCube and ANTARES measurements of neutrino flux from the Galactic Ridge 
are consistent with each other and with the Fermi-LAT measurement of the diffuse $\gamma$-ray flux from the Galactic Ridge at TeV energies. There are two important consequences of this fact. 

First, the contribution of leptonic sources and diffuse electron population to the diffuse gamma-ray flux is subdominant in the Galactic Ridge region. Thus, this region is the first confirmed multi-messenger hadronic source found in the Milky Way Galaxy.

Second, the common neutrino and gamma-ray spectrum has a slope of around 2.5 in a wide energy range from 10 GeV to 10-100 TeV. This is different from the locally measured cosmic ray spectrum, which has a power-law index of around 2.7. Thus, the cosmic ray spectrum is not universal in the Galaxy. This was conjectured based on the gamma-ray measurements \cite{2000A&A...362..937A,2015arXiv150507601N,2016PhRvD..93l3007Y}, but only the measurements of ANTARES and IceCube provide a definitive verification of the spatial variability of the slope of the Milky Way cosmic ray spectrum.

\section*{Acknowledgments} 
M.L. is a Postdoctoral Researcher of the Fonds de la Recherche Scientifique - FNRS.

\bibliography{refs.bib}
\end{document}